\documentstyle[twocolumn,prb,aps]{revtex}

\newcommand{\unite}[1]{\mbox{$\rm #1$}}

\newcommand{\Order}{{\cal O}}

\newcommand{\be}{\begin{equation}}
\newcommand{\ee}{\end{equation}}

\def\draftfig{1}

\begin{document}

\draft

\preprint{}

\title{Macroscopic Quantum Coherence in Ferrimagnets}

\author{Alain Chiolero and Daniel Loss}

\address{Department of Physics, University of Basel,\\
Klingelbergstrasse 82, 4056 Basel, Switzerland}

\date{\today}

\maketitle

\begin{abstract}

We study macroscopic quantum coherence (MQC) in small magnetic
particles where the magnetization (in ferromagnets) or the N\'eel
vector (in antiferromagnets) can tunnel between energy minima.  We
consider here the more general case of MQC in ferrimagnets by
studying a model for a mesoscopic antiferromagnet with an
uncompensated magnetic moment.  Through semi-classical calculations we show
that even a small moment has a drastic effect on MQC. In particular,
there is a rapid crossover to a regime where the MQC tunnel splitting
is equal to that obtained for a ferromagnet, even though the system is
still an antiferromagnet for all other aspects.  We calculate this
tunnel splitting via instanton methods and compare it with
numerical evaluations. As an application 
we re-examine the experimental evidence for MQC
in ferritin and show that even though
the  uncompensated  moment
of ferritin is small it greatly modifies the MQC behavior.  
The excess spin allows us to
extract values for
experimental parameters without making any assumption about the classical 
attempt frequency, in contrast to previous fits.
Finally, we also discuss the implications of our results for MQC in 
molecular magnets.

\end{abstract}



\narrowtext

\section{Introduction}

Quantum phenomena, like tunneling, are usually limited to the atomic
world. However, under some stringent conditions it is also possible
for a macroscopic variable to exhibit quantum behavior. Over the last
15 years, a lot of experimental and theoretical work has been devoted
to the quest of such systems. The clearest evidence so far has been
observed in Josephson junctions, where quantum tunneling of the
current phase can occur\cite{intro:Legget87,intro:Mart87}

More recently, another class of systems where macroscopic quantum
phenomena might be possible has been considered: small magnetic
particles \cite{intro:Chud88,intro:MQT}. In a grain too small to accommodate 
more than one magnetic domain,
all the spins are locked together by the exchange interaction, and only their
global orientation can change. Because of magnetic anisotropies and
external fields, the grain energy depends on this orientation. Two
type of phenomena can be envisaged. First, macroscopic quantum
tunneling (MQT), where the magnetization tunnels through an energy
barrier from one local minimum to the global one. Second, macroscopic
quantum coherence (MQC), where the classical energy has several equivalent
minima, and the magnetization oscillates back and forth between them.
This oscillation should show up in magnetic noise and susceptibility
spectra \cite{intro:Awsch92}.

Theoretical calculations based on simple models were first performed for
ferromagnets (FM)\cite{intro:Chud88,schilling,vanhemmen,korenblit}.
Surprisingly, they revealed that MQC was possible in systems
containing as much as $10^5$ moments.  This generated a lot of
interest.  Later extensions\cite{intro:Barb90,intro:Krive90} of these
calculations to antiferromagnets (AFM), where the N\'eel vector is the
tunneling entity, showed that MQC should show up at higher
temperatures and higher frequencies than in FM grains of similar size.
This makes AFM more interesting for experimental purposes.

The models used in early calculations\cite{intro:Barb90,intro:Krive90}
for AFM are based on the 
assumption that the two
sublattices had an exactly equal number of sites, meaning that the
grain had no net magnetic moment. However, this condition is not met
experimentally in finite-size magnets, since because of statistical
fluctuations a small imbalance in the two sublattices is expected. 
This was recognized early\cite{intro:Barb90} and deemed as a practical 
advantage. As the
uncompensated spin is locked together with the N\'eel vector, it should mirror
the MQC oscillations; the resulting magnetization oscillations are
easier to detect than those of the N\'eel vector. Although
addressed theoretically \cite{model:Loss92,intro:Loss93,model:Duan95}
the influence of
this uncompensated moment on MQC for a physical system 
has not been  studied yet, which motivated the present investigation.

To get an idea about the magnitude of this uncompensated moment, let us
consider several systems of practical interest. Ferritin is a
macromolecule of high biological importance, for which MQC evidence
has been reported\cite{intro:Awsch92,intro:Loss93,intro:Awsch92b,%
intro:Gid94,intro:Gid95,Sciencereply}.  It has a magnetic core
containing up to $4500$ ${\rm Fe}^{3+}$ atoms, whose spins are
antiferromagnetically coupled. Yet a ferritin molecule at full loading
has a magnetic moment corresponding to that of roughly $50$ iron
atoms. Clearly, this is small on the scale of the particle.  
Other systems of interest are molecular 
magnets\cite{intro:Gatt94,intro:Kahn93}. These 
molecules contain ligands binding a number of magnetic atoms (so far,
up to $20$), whose spins are strongly coupled. In addition,
macroscopic samples of perfectly identical molecules can be prepared.
This makes
them interesting for MQC, because the tunneling frequency scales
exponentially with the number of atoms. In these molecules, competing ferro- 
and antiferromagnetic interactions usually result in ferrimagnetic structures
with uncompensated moments that can be as large as, or even larger than, 
the sublattice magnetization.

In this paper we will investigate the influence of the uncompensated
moment on MQC in ferrimagnets. In the two systems we have just discussed, the
magnitude of this moment ranges from large (most molecular magnets)
to seemingly negligible (ferritin).  Nevertheless, we will show that
even in ferritin uncompensated spins drastically alter the MQC
results and dominate the tunneling dynamics.  The  paper is
structured in the following way.  In Section \ref{model}, we introduce
our model for a ferrimagnet: an antiferromagnet coupled to an
uncompensated moment.  We discuss its energy spectrum semi-classically
by looking at its instanton solutions in the regime of interest for
experiments.  In Section \ref{applic}, we apply our results to ferritin and
molecular magnets. We
conclude with a discussion in Section \ref{concl}.

\section{Model and semi-classical results}
\label{model}

In an antiferromagnetic grain much smaller than the typical domain
wall width, the N\'eel vector 
is expected to be uniform (Stoner-Wolfarth model). Its low-energy dynamics is
described by an effective action analogous to that of a rigid
rotator\cite{model:Andreev80,intro:Barb90,model:Frad91}. This model
was extended\cite{model:Loss92} to include the effects of an
uncompensated moment, which is treated as one (large) single spin
coupled to the AFM core. In the semi-classical regime, and using the
coherent-state path-integral representation\cite{model:Frad91,braunloss}, one
gets the following (Euclidean) action in the limit where the excess spin 
is strongly coupled to the N\'eel vector \cite{model:Loss92}
\begin{eqnarray}
S_E&=&V\int\!\!d\tau\Bigl\{
{\chi_\perp\over2\gamma^2}(\dot{\theta}^2+\dot{\phi}^2\sin^2\theta)
+K_y\sin^2\theta\sin^2\phi\nonumber\\
&&\qquad{}+K_z\cos^2\theta\Bigr\}+ i\hbar
S\int\!\!d\tau\,\dot{\phi}(1-\cos\theta){,}
\label{model:action}
\end{eqnarray}
where $\theta$ and $\phi$ are the spherical coordinates of the N\'eel
vector, $V$ is the volume of the grain, $\chi_\perp$ is its
perpendicular magnetic susceptibility, $K_z\geq K_y>0$ its magnetic
anisotropies\cite{model:Convention}, $\gamma=2\mu_B/\hbar$, and $S$ is
the magnitude of the excess spin. We have therefore included an easy
axis anisotropy along the $x$-axis, with the $xy$-plane taken as the
easy plane.

Notice that the term $i\hbar S\int\!d\tau\,\dot{\phi}$ in the action
only depends on the initial and final states, and hence does not
affect the classical equations of motion.
However\cite{model:Loss92,model:Delft92}, it associates a different
phase factor $\exp(\pm2i\pi S)$ to instantons and anti-instantons.
This makes the semi-classical predictions conform with Kramers theorem:
If $S$ is half-integer, the contributions of instantons and
anti-instantons exactly cancel each other, and the tunnel splitting
disappears.  If $S$ is integer, the contributions sum up coherently,
and tunneling occurs. We will suppress this topological term in the
following discussion of the classical solutions but will reinstate it
in the final expression for the tunnel splitting.

In the absence of an excess spin, i.e.\ for $S=0$, the instanton and
anti-instanton solutions are given by
\be
\phi(\tau)=\pm2\arctan(\exp(\omega_{\rm AFM}\tau)){,}\qquad
\theta(\tau)=\pi/2{,}
\ee
with
\be
\hbar\omega_{\rm AFM}=\mu_B\sqrt{8K_y/\chi_\perp}{.}
\label{model:omegaAFM}
\ee
These solutions have an action
\be
S_{\rm AFM}/\hbar=V\sqrt{2K_y\chi_\perp}/\mu_B{.}
\label{model:actionAFM}
\ee
We will be interested in the regime $K_z\gg K_y$. Loss {\em et al.}
\cite{model:Loss92} have investigated the effect of a small excess
spin, small in the sense that $\hbar S/S_{\rm AFM}\ll K_z/K_y$. Their
results are based on approximations that can be self-consistently
checked in this regime.  They found that the preceding expressions for
$\omega_{\rm AFM}$ and $S_{\rm AFM}$ remain correct provided one
replaces $\chi_\perp$ by an effective susceptibility $\chi_\perp^{\rm
eff}=\chi_\perp+(2/K_z)(S\mu_B/V)^2$.

These results were later confirmed by Duan and
Garg\cite{model:Duan95}, using an alternative model for the
uncompensated moment. Unlike Loss {\em et al.}, Duan and Garg use a
variational approach, which gives results for the entire parameter
space. The drawback of the variational apporach is that it is not 
possible to assess the accuracy
of its predictions, this being dependent on the ansatz quality.

We will now discuss the instanton solutions of the action
(\ref{model:action}) for a {\em large\/} excess spin, more precisely
for $\hbar S/S_{\rm AFM}\gg1$. We will perform a number of
approximations, and will check their self-consistency at the end.

In the absence of an excess spin, the characteristic time scale is
$\omega_{\rm AFM}^{-1}$.  To simplify the equations of motion, we make
a change of variables to a reduced time $\bar{\tau}=\omega_{\rm
AFM}\tau$. To get finite-action solutions, $\phi$ must be real, and
$\theta-\pi/2$ must be purely imaginary. Defining
$i\vartheta=\theta-\pi/2$, one gets for the action
\begin{eqnarray}
S_E&=&{S_{\rm AFM}\over2}\int\!\!d\bar{\tau}\,
\hbox{$\textstyle{1\over2}$}\Bigl\{
-\dot{\vartheta}^2+\cosh^2\vartheta(\dot{\phi}^2+\sin^2\phi)\\
&&\qquad{}-{K_z\over K_y}\sinh^2\vartheta- 4{\hbar S\over S_{\rm
AFM}}\dot{\phi}\sinh\vartheta\Bigr\}{,}
\end{eqnarray}
where overdots now mean derivatives with respect to $\bar{\tau}$.
Notice that all the physics is determined by only two dimensionless
quantities, $\hbar S/S_{\rm AFM}$ and $K_z/K_y$. The equations of
motion are
\begin{eqnarray}
\ddot{\phi}&=&\hbox{$\textstyle{1\over2}$}\sin2\phi+
{2\over\cosh\vartheta}{\hbar S\over S_{\rm AFM}}\dot{\vartheta}-
2\dot{\vartheta}\dot{\phi}\tanh\vartheta
\label{model:ddotphi}\\
\ddot{\vartheta}&=&
\hbox{$\textstyle{1\over2}$}\sinh2\vartheta\left(-\dot{\phi}^2+
{K_z\over K_y}-\sin^2\phi\right)+ 2{\hbar S\over S_{\rm
AFM}}\dot{\phi}\cosh\vartheta{,}
\label{model:ddottheta}
\end{eqnarray}
with the boundary conditions for an instanton solution being
$\phi(-\infty)=0$, $\phi(+\infty)=\pi$, $\vartheta(\pm\infty)=0$.

Recalling that we are interested in the regime $\hbar S/S_{\rm
AFM}\gg1$, $K_z/K_y\gg1$ we see that in Eqs.\ (\ref{model:ddotphi}) and
(\ref{model:ddottheta}) the second-order time-derivatives have much
smaller prefactors than the first-order ones. Therefore, we expect the
second-order derivatives to be negligible, except in possible boundary
layers, small regions usually located close to the boundaries (though
not always, see the discussion in Ref.\
\onlinecite{model:Kev81}). It turns out, however, that there are no boundary
layers for our choice of boundary conditions. If we neglect the
$\ddot{\vartheta}$ term in Eq.\ (\ref{model:ddottheta}), and assume
that $\dot{\phi}^2+\sin^2\phi\ll K_z/K_y$, we get
\be
\vartheta=-{\rm arcsinh}\left(2{K_y\over K_z}{\hbar S\over S_{\rm
AFM}}
\dot{\phi}\right){.}
\ee
We will see later that $(K_y/K_z)(\hbar S/S_{\rm AFM})\dot{\phi}\ll1$.  We
can therefore expand the hyperbolic arcsinus to first order.  Substituting
then $\dot{\vartheta}$ into Eq.\ (\ref{model:ddotphi}), neglecting
$\dot{\phi}\sinh\vartheta$ with respect to $\hbar S/S_{\rm AFM}$, and
expanding $\cosh\vartheta$ to leading order, we obtain
\be
\left(1+4{K_y\over K_z}\left({\hbar S\over S_{\rm
AFM}}\right)^2\right)
\ddot{\phi}=\hbox{$\textstyle{1\over2}$}\sin2\phi{.}
\ee
Thus, one gets exactly the same equation for $\phi(\tau)$ as in the absence of
the excess moment, except that now the instanton frequency is given by
\be
\omega=\left( {1+4{K_y\over K_z}\left({\hbar S\over S_{\rm
AFM}}\right)^2}
\right)^{-1/2}{.}
\ee
We see that the uncompensated moment slows down the transition between
the two energy minima (by a factor given precisely by $\omega$). It is
now straightforward to check the self-consistency of the
approximations that we used\cite{model:Justification}.

Moreover, it is easy to check that, within the same level of
approximation, $\omega$ also gives the frequency of the small
oscillations about each minimum and thus describes the lowest energy level
spacing (in the absence of the tunneling splitting) in the wells.

If we reinstate the full units, the instanton frequency in the
ferrimagnetic regime ($S\neq0$) becomes
\be
\omega_{\rm Ferri}={2\lambda V\over\hbar S}\sqrt{K_yK_z}{,}
\ee
with
\be
\lambda=\left( 1+{1\over4}{K_z\over K_y}\left(S_{\rm AFM}\over\hbar
S\right)^2
\right)^{-1/2}{.}
\ee
For the calculation of the action, since\cite{model:Justification}
$\vartheta$ is at most of order $\sqrt{K_y/K_z}$, we keep only the
lowest-order terms in the Taylor expansions of $\sinh\vartheta$ and
$\cosh\vartheta$. This gives
\be
S_{\rm Ferri}= {2\hbar S\over\lambda}\sqrt{K_y/K_z}\left\{1-
{1\over24}\lambda^4\left(S_{\rm AFM}\over\hbar S\right)^2\right\}{.}
\ee
Let us discuss these results in several limits. First, for an excess
spin becoming large, i.e.\ for $S\rightarrow\infty$, one sees that
$\lambda\rightarrow1$. Hence, $\omega_{\rm Ferri}$ and $S_{\rm Ferri}$
tend to the expressions expected for pure {\em
ferromagnets}\cite{schilling,vanhemmen,intro:Chud88} (with $K_z\gg K_y$),
\be
\omega_{\rm FM}={2V\over\hbar S}\sqrt{K_yK_z}{,}\qquad S_{\rm
FM}=2\hbar S\sqrt{K_y/K_z}{.}
\ee
However, the role of the total spin  in the ferromagnetic case is
now played by the excess spin $S$! To recover the FM regime starting from
an AFM might seem at first surprising, but should have been expected.
As the imbalance between the two sublattices grows, the dynamics
becomes that of a large spin coupled to a small AFM, i.e.\ a
ferromagnetic dynamics. This continuous transition between the seemingly
unrelated FM and AFM regimes provides a useful check of the model.

In the opposite limit of a small excess spin, i.e.\ for
$S\rightarrow0$, we would not expect our results to be accurate, since
they were derived under the assumption that $\hbar S/S_{\rm AFM}\gg1$.
However, it is easy to check that our expression for $\omega_{\rm
Ferri}$ is in fact equal to that derived in Ref.\
\onlinecite{model:Loss92}, where $\chi_\perp$ in Eq.\
(\ref{model:omegaAFM}) is replaced by an effective susceptibility
$\chi_\perp^{\rm eff}=\chi_\perp+(2/K_z)(S\mu_B/V)^2$.

Therefore, our results smoothly connect to the small excess spin
regime. Our analytical expressions are valid over the whole range of
$S$ (but for $K_z\gg K_y$), i.e.\ all the way from the AFM to the FM
regime. An alternative way to interpret them is to notice that in our
expression for $S_{\rm Ferri}$, the term $\lambda^4(S_{\rm AFM}/\hbar
S)^2$ lies between $0$ and $K_y/K_z$, and hence 
is a small
correction. The remaining term is exactly that expected for a pure
ferromagnet, but with an effective spin $S/\lambda$; in a sense, all
MQC splittings are given by the FM results.

The various regimes can be characterized by the value of $\lambda$. For a
small excess spin, such that $\lambda\approx2\sqrt{K_y/K_z}(\hbar
S/S_{\rm AFM})\approx0$, we are in an AFM regime where (to
first order) only $\chi_\perp$ and $K_y$ enter the action and the
instanton frequency. For a large excess spin, such that
$\lambda\approx1$, we are in a FM regime, where (again to first order)
only $K_y$ and $K_z$ are relevant. The crossover occurs for $\hbar
S/S_{\rm AFM}\approx\sqrt{K_z/K_y}$. We can get an estimate of the
critical value of the uncompensated moment by noting that, if MQC
is to be observable, $S_{\rm AFM}/\hbar$ should at most be of the
order of $10$. For a ratio $K_z/K_y$ of about $10$, the crossover would
occur for $S\approx30$, i.e.\ an uncompensated moment of $60\mu_B$.
This excess moment is quite small and a large enough system 
might be indistinguishable from a pure AFM for all its properties 
{\em except\/} for its MQC behavior.

To assess the accuracy of our analytical predictions, and in
particular to estimate how large the ratio $K_z/K_y$ needs to be, we
can compare them with direct numerical calculations of the instanton
solution of Eqs.\ (\ref{model:ddotphi}) and (\ref{model:ddottheta}).
Let us first fix the ratio $K_z/K_y$ to a value of $30$, and vary
$\hbar S/S_{\rm AFM}$.  We plot in Fig.\ \ref{model:Kz_Ky_30} the
instanton frequency\cite{model:inst_freq} and action.  As expected in
that regime, the analytical and numerical results compare extremely
well.  Trying now with a much smaller ratio $K_z/K_y=5$, where our
approximations are not so well justified, we get the results plotted
in Fig.\ \ref{model:Kz_Ky_5}.  Our analytical results for the
instanton frequency are accurate for small and large values of $\hbar
S/S_{\rm AFM}$; deviations are apparent (and expected) in the
intermediate regime. For the action, good results are obtained for
small $\hbar S/S_{\rm AFM}$; for large values the discrepancies are
due to our first-order expansion of $\sinh\theta$ and $\cosh\theta$ in
the action. 
We have verified that
inclusion of higher-order terms gives excellent agreement between
analytical and numerical results.

Alternatively, we can fix $\hbar S/S_{\rm AFM}$ and vary $K_z/K_y$.
The results for the choice $\hbar S/S_{\rm AFM}=10$ are shown in Fig.\
\ref{model:S_Safm10}.  Even though our analytical expressions were
derived under the assumption that $K_z\gg K_y$, we see that they give
accurate predictions down to $K_z/K_y\approx2$. For still lower values
the true action diverges, whereas our expressions remain finite. This
divergence reflects the appearance of a new symmetry: For $K_y=K_z$,
$S_x$ is a conserved quantity, and consequently tunneling is blocked
if $S\neq0$. The tunnel splitting must therefore vanish.

With our solution for the instanton, and after evaluation of the
fluctuation determinant (again for $K_z\gg K_y$), we get the following
prediction for the tunnel splitting
\be
\Delta_0=8\hbar\omega_{\rm Ferri}\sqrt{S_{\rm Ferri}\over2\pi\hbar}
\left|\cos(\pi S)\right|
e^{-S_{\rm Ferri}/\hbar}{.}
\label{model:splitting}
\ee
The term $|\cos(\pi S)|$ is due\cite{model:Loss92} to
the topological term, which we have reinstated.

Using the criterion that the MQC resonance should
show up at temperatures low enough that the quantum transition rates
between the degenerate energy minima are higher than the classical rates, 
one gets for the
crossover temperature
\be
k_BT^*=K_yV\hbar/S_{\rm Ferri}{.}
\label{model:omega}
\ee
Up to relative corrections of order $K_y/K_z$, $S_{\rm Ferri}$ is
equal to $4K_yV/\omega_{\rm Ferri}$, and therefore
$k_BT^*\approx\hbar\omega_{\rm Ferri}/4$. Since the two lowest levels
are separated from the others by an energy $\hbar\omega_{\rm Ferri}$,
the criterion means that the transition to the quantum regime occurs
for temperatures such that only the lowest two levels are populated.

To conclude this section, let us come back to the issue of the best
system to detect MQC, ferromagnets or antiferromagnets.  Early
calculations of the tunnel splitting in (pure) FM and AFM gave
$\Delta_{\rm FM}\sim \omega_0\exp(-Ns\sqrt{K_y/K_z})$ and $\Delta_{\rm
AFM}\sim \omega_0\exp(-Ns\sqrt{K_y/J})$, where $N$ is the number of
sites, $s$ is the magnitude of the spins, $J$ is the exchange
coupling, and $\omega_0$ a characteristic frequency.  Since $J\gg
K_z$, MQC should be easier to observe in AFM than in FM. Our results
for an AFM with an uncompensated moment give $\Delta_{\rm Ferri}\sim
\omega_0\exp(-N_es\sqrt{K_y/K_z})$, where $N_e$ is the number of 
excess spins.  Even though this is similar in structure to the FM
result, AFM nevertheless remain more favorable to observe MQC since
typically $N_e\ll N$.

\section{Applications}
\label{applic}

\subsection{Ferritin}

Ferritin is a macromolecule of high biological importance, naturally
occurring in many living organisms where it is responsible for iron
transport. It has a chestnut structure, with an outside shell about
$2\,{\rm nm}$ thick enclosing a core of iron oxide. The roughly
spherical core has a diameter of $7.5\,{\rm nm}$, and can contain a
variable number of irons, up to a maximum of $4500$. The oxide
appears\cite{applic:Ford84} structurally similar to
ferrihydrite\cite{applic:Towe67,applic:Chuck,applic:contro} .

Ferritin molecules have a small magnetic moment.
Measurements\cite{intro:Awsch92,applic:Kil95} on fully loaded grains
of horse-spleen ferritin give values ranging between $217\mu_B$ and
$316\mu_B$. This range corresponds to roughly $45$--$60$ parallel iron
spins, and, if compared to the moment expected if all the spins were
parallel (about $4500\times5\mu_B$), provides evidence for an
antiferromagnetic core.

MQC evidence has been reported\cite{intro:Awsch92,intro:Awsch92b,%
intro:Loss93,intro:Gid94,intro:Gid95} in diluted samples of
horse-spleen ferritin.  Since the resonance frequency depends
exponentially on the amount of AFM material, it is crucial to be able
to select particles with a well-defined loading. For that reason, the
first experiments\cite{intro:Awsch92,intro:Awsch92b,intro:Loss93} used
fully loaded grains, which could be reliably sifted out. In both the
magnetic noise and susceptibility spectra, a peak at $\nu_{\rm
res}=940\,{\rm kHz}$ appeared below $T^*=200\,{\rm mK}$, and was fully
developed at about $30\,{\rm mK}$. From its field- and
concentration-dependence, and from other cross-checks, this resonance
was attributed to MQC \cite{intro:Awsch92}. Later 
experiments\cite{intro:Gid94,intro:Gid95}
exploited recent advances in synthesis of ferritin to artificially
engineer grains with a given loading.  The resonance frequency was
found to decrease exponentially with the loading, as expected from the
MQC predictions for a (pure) AFM.

The previous quantitative analysis of the experimental
results\cite{intro:Awsch92,intro:Awsch92b,intro:Loss93,%
intro:Gid94,intro:Gid95} used the following estimate for the MQC
resonance frequency, $\omega_{\rm res}=\omega_0\exp(-S_{\rm
AFM}/\hbar)$, with $\omega_0$ {\em fixed\/} to a value of
$10^{10}\,\unite{s}^{-1}$ (and not measured or calculated from the
inferred $\chi_\perp$ and $K_y$).  The crossover temperature was
estimated as the one where the rate of thermal fluctuations between
the two minima is equal to the MQC rate, giving\cite{intro:Barb90}
$T^*=\mu_B\sqrt{K_y/2\chi_\perp}/k_B$.  From the experimental results
for $\omega_{\rm res}$ and $T^*$, the following values were derived
\be
\chi_\perp=5.2\cdot10^{-5}\,{\unite{emu}\over\unite{G}\,\unite{cm^3}}{,}
\qquad K_y=0.95\cdot10^3\,{\unite{erg}\over\unite{cm^3}}{.}
\label{applic:param_old}
\ee
An assumed value for $\omega_0$ had to be used for lack of experimental
data at that time. Notice that $\omega_0/2\pi$ gives the
frequency of the small oscillations around the minima.
In Kramers approach, it should therefore approximately equal
the classical attempt frequency $f_0$. Recent
measurements\cite{applic:Kil95,applic:Dickson93} of $f_0$ in
fully loaded grains of horse-spleen ferritin gave much larger values,
ranging between $10^{11}$ and $10^{12}\,\unite{s}^{-1}$.
The spread in these results is indicative of the difficulty of these
measurements, but they both point to values of $\omega_0$ much
larger than $10^{10}\,\unite{s}^{-1}$.

This would hardly matter if MQC measurements had been performed on
fully loaded grains {\em only}. Slightly different values for the
experimental parameters $\chi_\perp$ and $K_y$ would result from the
fit.  However, $\omega_0$ plays an important role in the scaling 
of the resonance
frequency with the grain loading, and within the pure AFM approach,
where $\omega_{\rm res}=\omega_0\exp(-S_{\rm AFM}/\hbar)$, it is
readily seen that a value of $10^{10}\,\unite{s}^{-1}$ is necessary to
reproduce the experimental data. Since this value is far too small compared
with the independently measured \cite{applic:Kil95,applic:Dickson93} one,
one is therefore faced with a consistency problem, which we shall
address now.

Being so small on the scale of the molecule, the uncompensated moment
was neglected in the previous quantitative analysis. That it should play
an important role can be easily appreciated. If one takes at face
value the parameters (\ref{applic:param_old}), one gets for the action
a value $S_{\rm AFM}/\hbar\approx7.5$. With a measured excess spin
$S_0\approx100$--$150$, the ratio $\hbar S/S_{\rm AFM}$ would be in
the range $13$--$20$. Thus, it is not possible to neglect the
uncompensated moment.

However, the following problem now arises. Experimentally, two quantities 
were measured in
the MQC experiments: the resonance frequency $\nu_{\rm res}$ and the
crossover temperature $T^*$.  Without excess spin, these quantities
depend on two parameters only ($K_y$ and $\chi_\perp$), which are
therefore unambiguously fixed. With an excess spin, three parameters
are relevant ($K_y$, $K_z$, and $\chi_\perp$), and the MQC experimental
results alone are not sufficient to determine them.

We can circumvent this problem by using an additional parameter, namely the
N\'eel temperature $T_N$.  We use the simple
estimate\cite{intro:Barb90,applic:Bar85}
$\chi_\perp=\mu_B^2N/k_BT_NV$, where $N$ is the number of spins (about
$4500$ for a fully loaded grain), and $V$ is the core volume (which we
estimate as that of a sphere of $7.5\,{\rm nm}$ in diameter).  From
the value $T_N=240\,\unite{K}$ reported for
ferritin\cite{applic:Baum89}, one gets $\chi_\perp=5\cdot10^{-5}\,
{\unite{emu}/\unite{G}\,\unite{cm}^3}$.

If we suppose that the excess spin is $S_0=100$, Eqs.\
(\ref{model:splitting}) and (\ref{model:omega}) give
\be
K_y=1.5\cdot10^3\,{\unite{erg}\over\unite{cm^3}}{,}\qquad
K_z=9.8\cdot10^5\,{\unite{erg}\over\unite{cm^3}}\,\,{.}
\label{model:param_new}
\ee
The value for $K_y$ is a bit larger than the one obtained neglecting
the uncompensated moment; $K_z$ was not determined in the original
fit. These results depend on the values taken for $S_0$ and
$\chi_\perp$. For example, repeating the fits with $S_0=150$ we find that
$K_z=2.2\cdot10^6\,\unite{erg}/\unite{cm^3}$, whereas with
$\chi_\perp=10^{-5}\, {\unite{emu}/\unite{G}\,\unite{cm}^3}$, $K_z$
reduces to $4.7\cdot10^5\,\unite{erg}/\unite{cm^3}$, while
$K_y$ is left unchanged.

{}From the parameters (\ref{model:param_new}), one gets for a fully
loaded grain $\omega_{\rm Ferri}=10^{11}\,{\rm s}^{-1}$, $S_{\rm
Ferri}\approx12\hbar$, and $\lambda\approx0.65$. We are therefore not
quite in the FM regime, which explains why the choice of $\chi_\perp$
still has an influence. The value of $\omega_{\rm Ferri}/2\pi$ is a bit
lower than the experimental
results\cite{applic:Kil95,applic:Dickson93} for the classical attempt
frequency $f_0$ (between $10^{11}$ and
$10^{12}\,\unite{s}^{-1}$), but not unreasonably so considering
the simplicity of the model and the experimental uncertainties.

To get predictions for the volume dependence of the MQC resonance
frequency, we need first to know how the excess spin depends on the
grain loading. To the best of our knowledge, no experimental study
exists. We will postulate a simple scaling law, $S=S_0(V/V_0)^\nu$,
where $S$ and $V$ are the excess spin and volume of a grain, $S_0$ and
$V_0$ corresponding to a fully loaded grain. For the exponent $\nu$,
several models have been proposed. Suppose the uncompensated moment is
solely due to a difference in the number of A and B sites at the
surface. Barbara and Chudnovsky\cite{intro:Barb90} have argued that,
since the number of surface spins scales like $V^{2/3}$, the excess
spin (being the difference of two such random variables) should
therefore scale like $V^{1/3}$.  N\'eel has proposed several other
models\cite{applic:Neel61}, giving a scaling of $S$ with either
$V^{1/3}$, $V^{1/2}$, or $V^{2/3}$.  The theoretical picture is therefore
far from settled and in the absence of experimental data we will
consider all three scaling laws.

The predictions for the MQC frequency as a function of the number of
spins are shown in Fig.\ \ref{applic:scaling}, together with the
experimental measurements. A scaling of $S$ with $V^{1/3}$ clearly
gives the best results, whereas a scaling with $V^{1/2}$ predicts 
frequencies that
are too large at low loadings. As this gets even worse with a
$V^{2/3}$-scaling, we have not displayed these results in Fig.\
\ref{applic:scaling} to avoid cluttering.

Notice that the scaling is more complicated than in the simple AFM
picture used in the previous analysis of the experiments. In the latter
case, $\omega_{\rm res}=\omega_0\exp(-S_{\rm AFM}/\hbar)$, with
$S_{\rm AFM}$ being linear in the total number of spins. Hence, the resonance
frequency depends exponentially on the loading. This is no more the
case in the ferrimagnetic case. In the FM ($\lambda\approx1$) and AFM
($\lambda\approx0$) limits, the frequency does indeed depend
exponentially on the loading, though in a different way in the two
regimes. In the intermediate regime, there is a smooth crossover
between the two scalings.

The value taken for the uncompensated spin $S_0$ has nearly no effect
on the scaling predictions. (We did not show in Fig.\
\ref{applic:scaling} the results obtained with $S_0=150$, the curves
being indistinguishable.)  On the other hand, the value taken for
$\chi_\perp$ affects the scaling markedly. We show in Fig.\
\ref{applic:scaling} the predictions obtained with
$\chi_\perp=10^{-5}\, {\unite{emu}/\unite{G}\,\unite{cm}^3}$, which
suggests an alternative approach.  We recall that we have estimated
$\chi_\perp$ from the measured N\'eel temperature.  Instead, we could
have determined $\chi_\perp$ from the scaling measurements. In this way, all 
the
parameters ($\chi_\perp$, $K_y$, and $K_z$) would have been extracted
from MQC measurements performed in the same regime of temperature,
concentration, and field.  However, this would have required that we
know beforehand the scaling of the excess spin (either from more
sophisticated models, or from independent measurements).  In the
absence of such data, the use of the N\'eel temperature seems more
reliable.

As we have indicated earlier, $\lambda$ is of the order of $0.65$ for
fully loaded grains, which are therefore only close to the FM regime.
However, $\lambda$ increases steadily with diminishing loading,
reaching a value of about $0.92$ for a grain with $1000$ irons.
Because of the magnitude of the ratio $K_z/K_y$, it turns out that,
despite their quite large uncompensated moment, fully loaded grains
are still within the range of validity of the small excess spin
approach of Ref.\ \onlinecite{model:Loss92}. However, this is no more
true for partially loaded grains.

One last remark about expression (\ref{model:splitting}) for the
tunnel splitting. Strictly speaking, we see that it predicts a
vanishing splitting in the limit of small volumes. However, no hint of
this appears in Fig. \ref{applic:scaling}, which rather suggests that
the splitting continues to increase. In fact, it is only below extremely
small sizes (grain with a few tens of spins) that a sudden downturn
occurs. Such sizes are not shown in Fig.\
\ref{applic:scaling}. We note, however, that this is not a real effect, as the
level splitting of small clusters has no reason to vanish, but rather
signals a breakdown of the instanton approach\cite{applic:breakdown}.

\subsection{Molecular Magnets}

As we mentioned in the Introduction, most molecular magnets are
ferrimagnets. For example\cite{intro:Gatt94}, the magnetic properties
of ${\rm Mn}_{12}{\rm O}_{12}\hbox{(acetate)}_{16}$ point to a
structure where the spins of $8$ ${\rm Mn}^{3+}$ ($S=2$) are parallel
to each other and antiparallel to those of $4$ ${\rm Mn}^{4+}$
($S=3/2$). One can model this system as an AFM having two sublattices
carrying each a spin $S=6$, with an uncompensated spin $S=10$ coupled
to its N\'eel vector. Likewise\cite{intro:Gatt94}, $[{\rm
Mn}\hbox{(hfac)}_2\hbox{NITPh}]_6$, with $6$ ${\rm Mn}^{2+}$ ($S=5/2$)
parallel to each other and antiparallel to $6$ $\hbox{NITPh}$
($S=1/2$), can be viewed as an AFM with sublattice spin $S=3$ and
excess spin $S=12$. Finally\cite{intro:Gatt94}, $[{\rm
V}_{15}\hbox{As}_6{\rm O}_{42}({\rm H}_2{\rm O})]^{-}$ has $15$ spins
$1/2$ antiferromagnetically coupled, which in the AFM model gives a
sublattice spin of $S=7/2$ and an excess spin $S=1/2$.

To discuss the influence of the excess spin on the tunnel splitting in
molecular magnets, it will prove
convenient to express our results in terms of microscopic quantities.
For this, we consider a perfect AFM containing $N$ spins $s$ (with coordination
number $z$), whose N\'eel vector is coupled to an uncompensated
moment formed by $N_e$ spins $s$. Our results read (again for $k_y\ll k_z$)
\be
S_{\rm Ferri}={2\hbar N_es\over\lambda}\sqrt{k_y/k_z}\left\{ 1-
{1\over24}\lambda^4\left(S_{\rm AFM}\over\hbar S\right)^2\right\}{,}
\ee
with
\be
S_{\rm AFM}=2Ns\hbar\sqrt{k_y/Jz}{,}
\ee
\be
{\hbar S\over S_{\rm AFM}}={N_e\over2N\sqrt{k_y/Jz}}{,}
\ee
\be
\lambda=\left(1+{k_z\over Jz}\left(N\over
N_e\right)^2\right)^{-1/2}{,}
\ee
where $J$ is the exchange coupling, $k_y$ and $k_z$ are the local
anisotropies (felt by all spins). 
We are not aware of any molecular magnet for which all the magnetic 
anisotropies have been
measured, hence we cannot give predictions for the tunnel splitting.
However, many magnets have an excess spin of about the same order of magnitude
as the sublattice spin, and since we expect $k_y$ and $k_z$ to be 
much smaller than $J$, this
implies that $\hbar S/S_{\rm AFM}\gg1$ and that $\lambda\approx1$.
Therefore, they should be fully in the ferromagnetic regime. With respect to
the results that would be obtained by neglecting the excess spin,
notice that only the tunnel splitting is drastically different; the
crossover temperature to the quantum regime will remain of the same
order of magnitude.

\section{Discussion}
\label{concl}

We have studied macroscopic quantum coherence in a simple model for a
ferrimagnet, taken as an antiferromagnet whose N\'eel vector is
coupled to a large spin mimicking the uncompensated moment.

We have obtained approximate instanton solutions for $\hbar S/S_{\rm
AFM}$ large and $K_z\gg K_y$, where $S$ is the excess spin, $S_{\rm
AFM}$ is the action of the instanton solutions in the absence of the
excess spin, and $K_z\geq K_y>0$ are the magnetic anisotropies. The
expressions derived in this regime turn out to be also valid for
$\hbar S/S_{\rm AFM}$ small. Hence, our results cover all cases from
FM to AFM. Known results are correctly recovered in both limits.
Numerical simulations fully confirm the validity of our
approximations, and this even for ratios $K_z/K_y$ as small as $2$.
Using these solutions, we have calculated the MQC tunnel splitting by
instanton methods, including the contribution from the
fluctuation-determinant.

Our results exhibit two surprising features. Intuitively, provided the
uncompensated moment is much smaller than the sublattice magnetization
(the tunneling entity), one would expect only small modifications of
the MQC resonance frequency. This criterion can be written as $S\ll Ns$,
where $S$ is the excess spin, $N$ is the number of sites, and $s$ is
the local spin. In fact, we have shown that the correct criterion is
much more drastic: The AFM regime requires that $S\ll Ns\sqrt{K_z/J}$,
where $J$ is the exchange coupling.  Since typically $J$ is of the
order of $10^9\,{\rm erg}/{\rm cm}^3$ and $K_z$ of the order of
$10^5\,{\rm erg}/{\rm cm}^3$, there is a large reduction factor over
the intuitive criterion. Even more surprisingly, for $S\gg
Ns\sqrt{K_z/J}$, one enters a FM regime, where the tunnel splitting is
given by the expressions for a FM, but with the excess spin taking the
place of the total spin.

The quantity $Ns\sqrt{K_z/J}$ determines the critical magnitude of the
excess spin beyond which one reaches the FM regime. We note that it can
be rewritten as $S_{\rm AFM}\sqrt{K_z/K_y}/\hbar$. For MQC to be
observable, $S_{\rm AFM}/\hbar$ cannot be much larger than $10$.
Hence, the critical spin is quite small. This puts strong limitations
on the possibility of observing pure AFM tunneling in mesoscopic grains.
Ultimately, this
shows that MQC is a very sensitive phenomenon and that a few additional spins
can modify the tunneling frequency markedly.

We have considered two applications. First, we have re-examined the
experimental evidence for MQC in ferritin.  Even though the magnetic
moment of this molecule is small, it modifies the MQC resonance
frequency already in the fully loaded grains used in the first
experiments \cite{intro:Awsch92}, and has even stronger effects in 
the partially loaded
grains studied later \cite{intro:Gid95}. Our theory has 
several advantages over the
original interpretation of the experiments. It does not make any
hypothesis on the classical attempt frequency $\omega_0$; instead, it
predicts a value in fair agreement with recent measurements. Moreover,
it reproduces the scaling of the resonance frequency as a function of
the loading, whereas the original theory would not unless one was to
arbitrarily set $\omega_0$ to a value much below the measured one.

The value for the magnetic anisotropy extracted from the original MQC
experiments was found to be much smaller than the one derived from
M\"ossbauer\cite{concl:Bell84} and SQUID\cite{concl:Teja94}
measurements of the blocking temperature in the superparamagnetic
regime. This was tentatively imputed\cite{intro:Gid94} to differences
in the fields and temperatures at which the various quantities are
measured. With respect to the original interpretation, our theory
gives a slightly larger value for $K_y$, yet still much smaller than
the one expected from the blocking temperature.

At that stage, further testing of our theory requires additional data,
either an independent measurement of the anisotropies in the MQC
regime, or a measurement of the magnetic moment of the ferritin
particles as a function of loading.

The second application we have considered is to  molecular magnets.
Systems like these are usually neither
pure FM nor AFM. Moreover, their small size renders them even more
sensitive to an excess spin.  Thus, our results should be useful for a
quantitative understanding of MQC in such systems.

\acknowledgments
We are grateful to D.D.\ Awschalom, H.-B.\ Braun, 
and J.\ Kyriakidis for useful
discussions. One of us (D.L.) also acknowledges support from
the US NSF PHY $94$-$07194$
and the hospitality of the ITP (Santa Barbara) where part of this work
has been performed.
This work has been supported by the NSERC of Canada 
and the Swiss NSF.

\begin{figure}

\ifnum\draftfig=1
  \vspace*{0cm}
\else
\begin{center}
\BoxedEPSF{Graphs/vs_S1.ps}
\end{center}
\fi

\caption{Instanton frequency and action as functions of 
$\hbar S/S_{\rm AFM}$ for $K_z/K_y=30$: numerical (diamonds) and
analytical (line) results.}
\label{model:Kz_Ky_30}

\end{figure}

\begin{figure}

\ifnum\draftfig=1
  \vspace*{0cm}
\else
\begin{center}
\BoxedEPSF{Graphs/vs_S2.ps}
\end{center}
\fi

\caption{Instanton frequency and action as functions of 
$\hbar S/S_{\rm AFM}$ for $K_z/K_y=5$: numerical (diamonds) and
analytical (line) results.}
\label{model:Kz_Ky_5}

\end{figure}

\begin{figure}

\ifnum\draftfig=1
  \vspace*{0cm}
\else
\begin{center}
\BoxedEPSF{Graphs/vs_K.ps}
\end{center}
\fi

\caption{Instanton frequency and action as functions of $K_z/K_y$ for
$\hbar S/S_{\rm AFM}=10$: numerical (diamonds) and
analytical (line) results.}
\label{model:S_Safm10}

\end{figure}

\begin{figure}

\ifnum\draftfig=1
  \vspace*{0cm}
\else
\begin{center}
\BoxedEPSF{Graphs/scaling.ps}
\end{center}
\fi

\caption{MQC frequency as a function of the grain loading: experimental
measurements (diamonds), and theoretical predictions obtained with
$\chi_\perp=5\cdot10^{-5}\, {\unite{emu}/\unite{G}\,\unite{cm}^3}$ for
a scaling exponent $\nu=1/3$ (continuous line), $\nu=1/2$ (dashed
line), and with $\chi_\perp=10^{-5}\, {\unite{emu}/\unite{G}\,
\unite{cm}^3}$, $\nu=1/3$ (long dashed line).}
\label{applic:scaling}

\end{figure}


\begin{references}
\bibitem{model:Convention} Notice that we have a factor $1/2$ of
difference in our definitions of $K_y$ and $K_z$ with respect to the
convention used in Ref.\ \onlinecite{intro:Barb90}.
\bibitem{model:inst_freq} The instanton frequency is defined as
$\phi'(\tau_0)$, with $\tau_0$ such that $\phi(\tau_0)=\pi/2$.
\bibitem{applic:breakdown} Instanton calculations are based
on the so-called dilute instanton-gas approximation, which requires
that the classical action be much larger than $\hbar$.
\end{references}
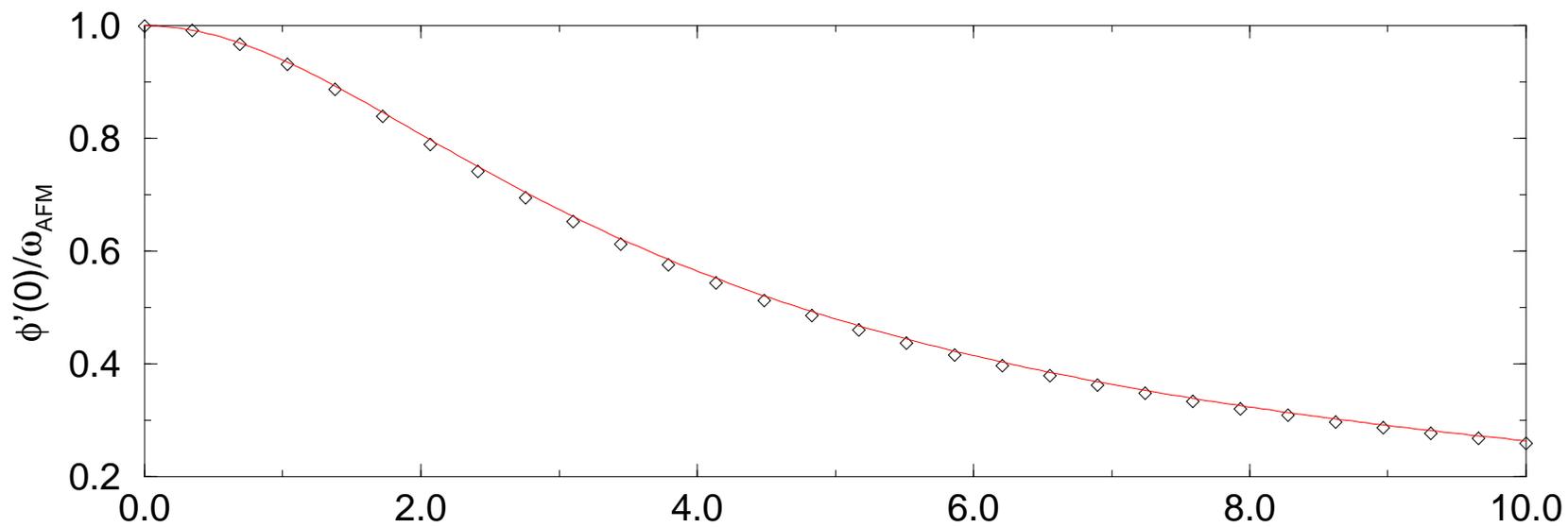
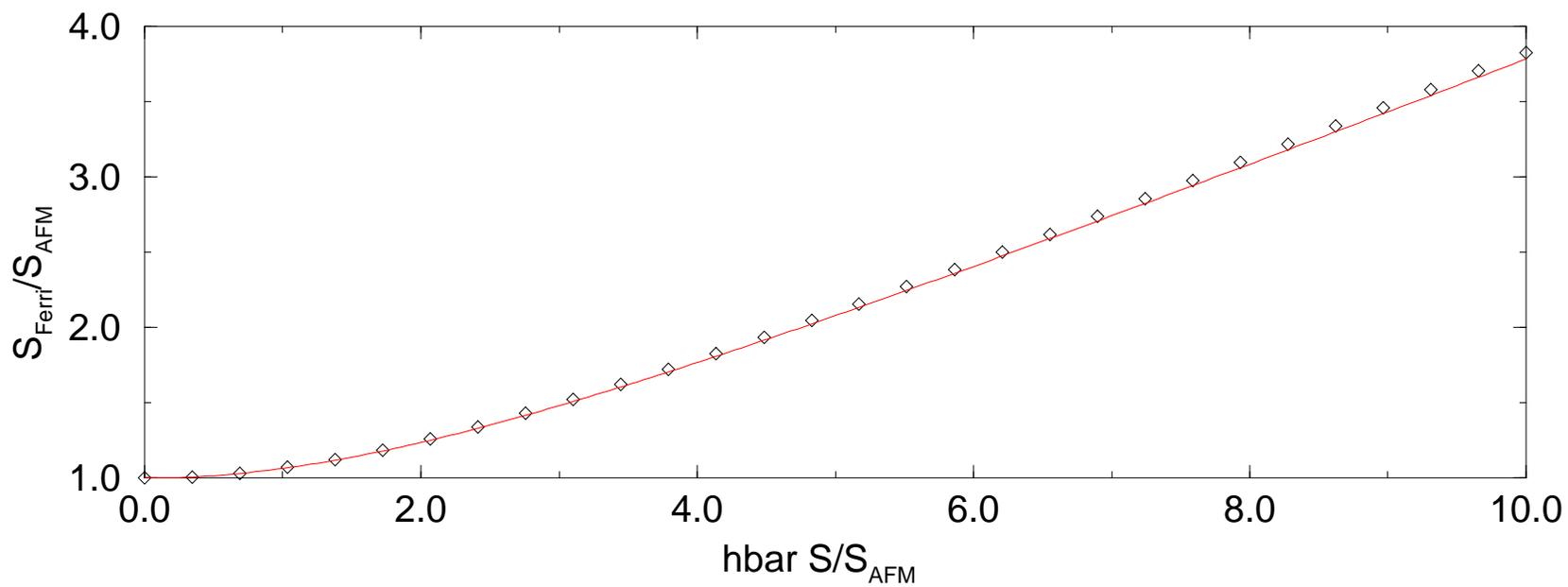
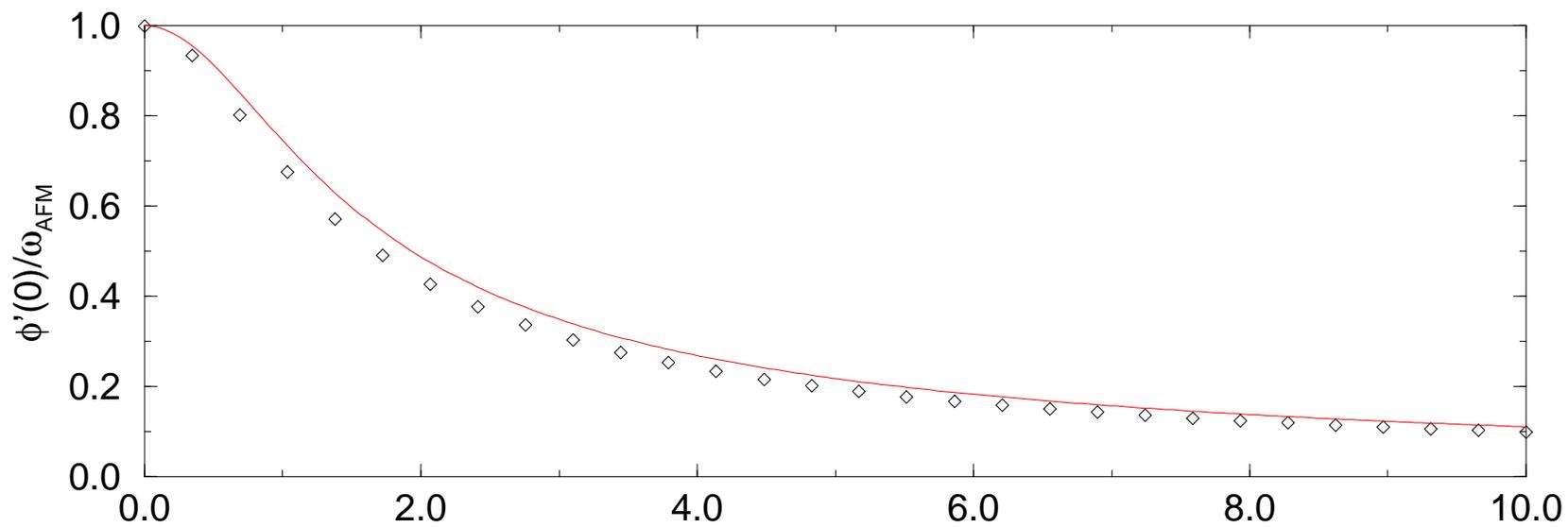
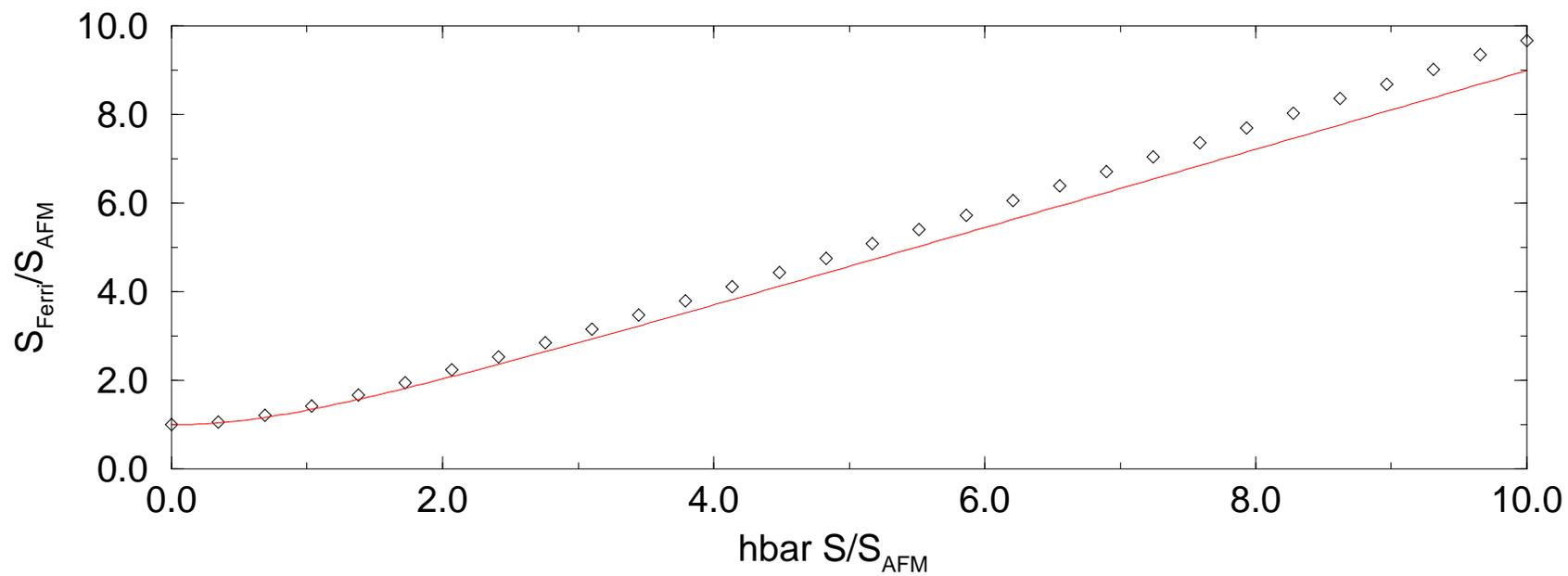
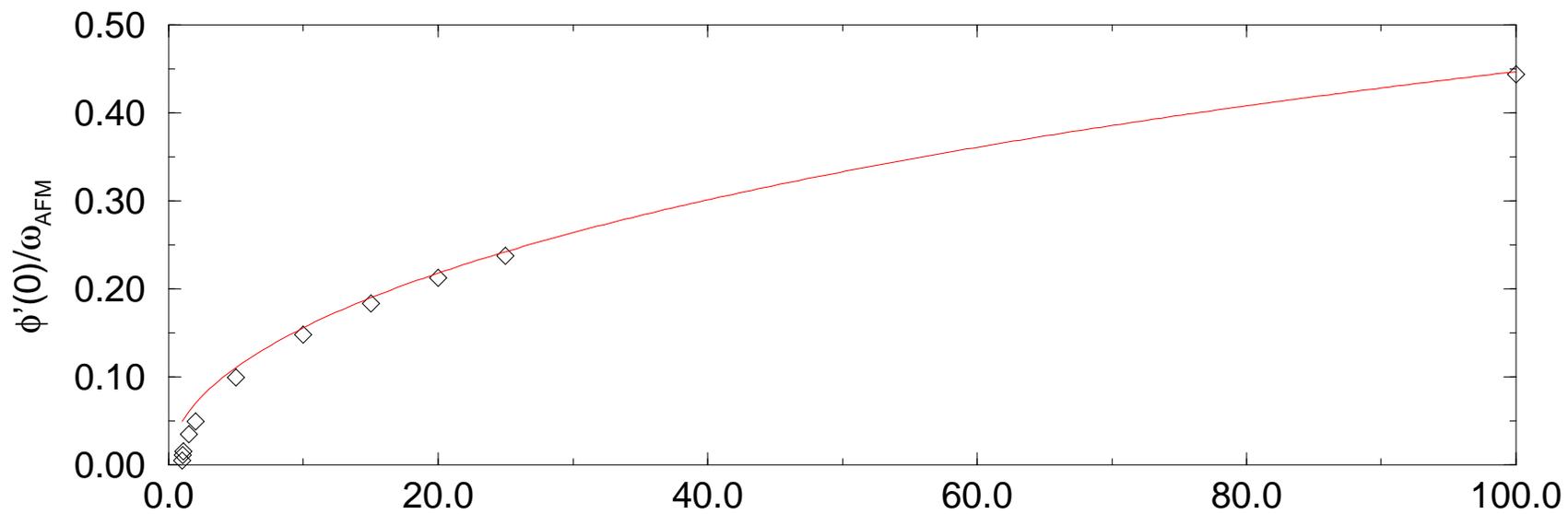
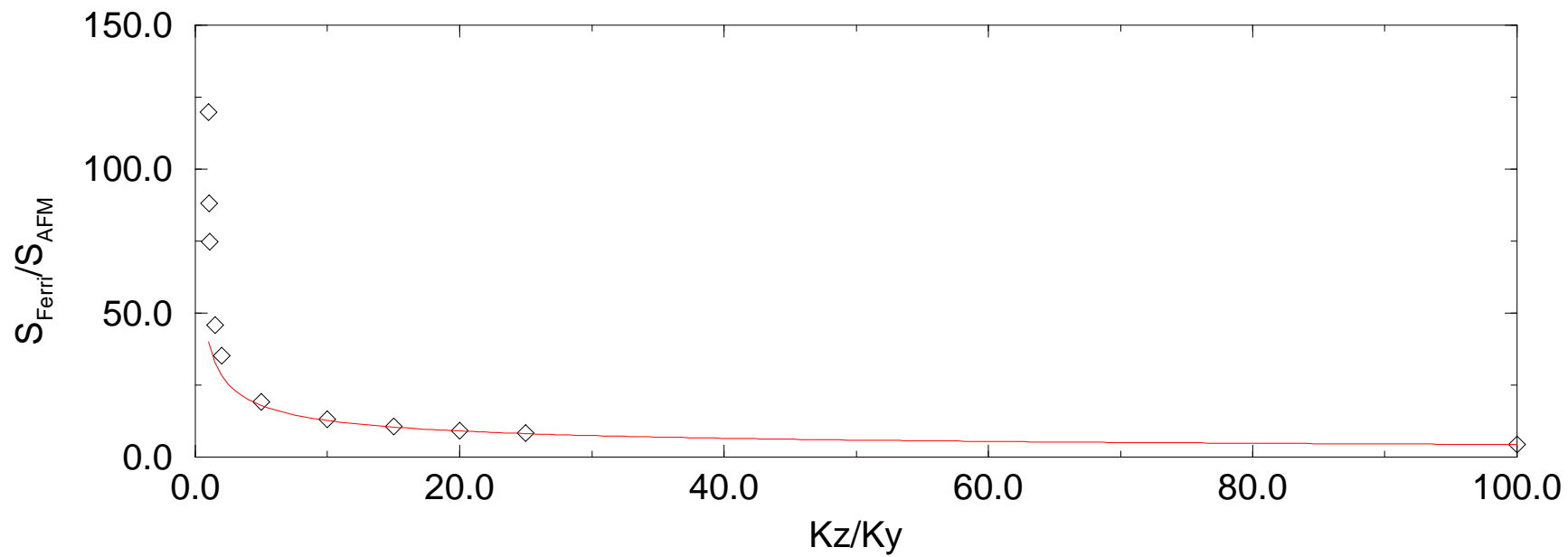
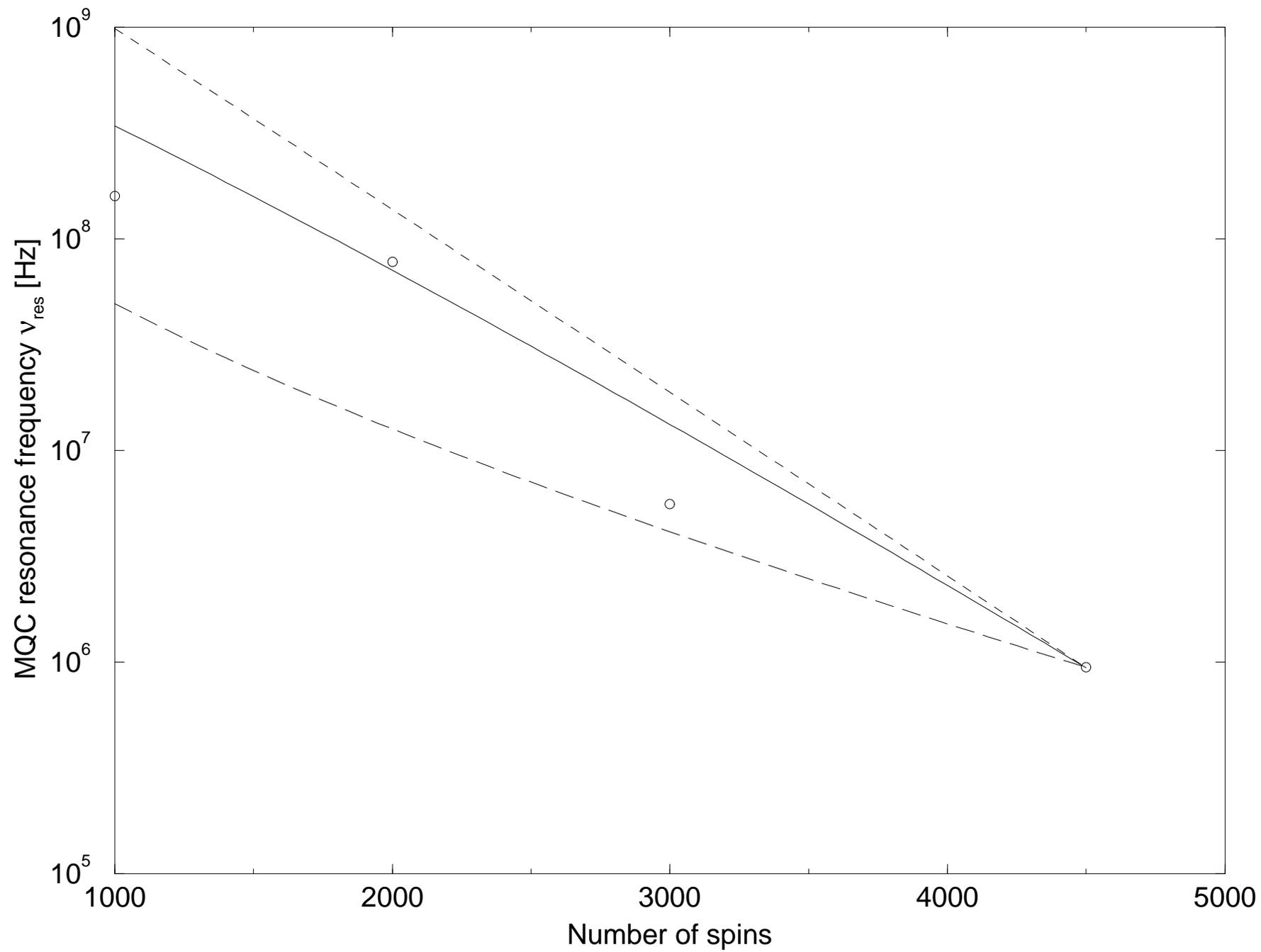
\end{document}